\documentclass[conference,letterpaper,10pt]{IEEEtran}
\usepackage{float}
\usepackage{cite}
\usepackage{amsmath,amssymb,amsfonts}
\usepackage{graphicx}
\usepackage{url}
\usepackage{flafter}
\usepackage{placeins}
\usepackage{algorithm}
\usepackage{algpseudocode}
\usepackage{array,tabularx}
\usepackage{microtype}
\algrenewcommand{\algorithmicrequire}{\textbf{Input:}}
\graphicspath{{figures/}{./}}
\makeatletter
\g@addto@macro\normalsize{\setlength{\abovedisplayskip}{5pt plus 2pt minus 1pt}\setlength{\belowdisplayskip}{5pt plus 2pt minus 1pt}\setlength{\belowdisplayshortskip}{5pt plus 2pt minus 1pt}}
\makeatother
\begin{document}
\title{Data Center-Aware Load Pickup in Power System Restoration}

\author{
\IEEEauthorblockN{
Yousef Abudyak, Saleh Sadeghi Gougheri, Mohsen Alizadeh, and Wei Sun
}
\IEEEauthorblockA{
\textit{Department of Electrical and Computer Engineering, University of Central Florida, Orlando, FL, USA}\\
\{yousefhusamyousef.abudyak, saleh.sadeghi, mohsen.alizadeh, sun\}@ucf.edu
}
}

\maketitle

\begin{abstract}
Data centers requiring continuous operation rely on onsite generation and battery storage during blackouts. Their return to grid supply must be coordinated with the restoration of other loads because demand fluctuations can affect system frequency. This paper adds constraints to a restoration optimization problem that coordinates load pickup, grid supply to data centers, onsite generation, and battery operation. These constraints reserve battery power and energy to accommodate bounded demand changes while maintaining approved grid supply to data centers. The method accounts for processing delays and checks resource limits using updated operating states before applying commands. Phasor-domain simulations of a low-inertia restoration scenario show that the method serves 760~MW of load outside the data centers and satisfies all tested operating limits. Nominal and frequency-aware benchmarks serve 840~MW and 800~MW, respectively, but violate operating limits. 
\end{abstract}


\section{Introduction}
Data centers requiring continuous operation rely on onsite generation and storage during blackouts \cite{wang_black-start_2026}. Returning to grid supply transfers their demand to a partially restored system alongside native load pickup, defined here as the restoration of loads outside the data centers. AI training and checkpoint cycles produce rapid demand changes \cite{nerc_large_loads_2025}. Coordinated onsite resources can contain these fluctuations \cite{omri_ai_2026}, but limited power and energy reserves can expose the recovering grid to frequency excursions, rate-of-change-of-frequency (RoCoF) violations, and voltage deviations.

Prior work addresses parts of this problem. Frequency-constrained restoration embeds nadir bounds in sequential mixed-integer linear programs (MILPs) \cite{zou_frequency_2025} and coordinates renewables and grid-forming (GFM) loads under nadir, RoCoF, and storage limits \cite{huang_integrating_2025}. Data-center studies coordinate workloads and emergency supplies \cite{wang_black-start_2026}, calibrate dynamic load models \cite{lu_dynamic_2026}, and assess reconnection timing and ramping from grid-side trajectories \cite{chao_dynamic_2026}. On the storage side, our earlier GFM battery energy storage system (BESS) study \cite{abudyak_mitigating_2026} demonstrated power tracking, voltage support, and islanding. Azizi \emph{et al.} \cite{azizi_strengthening_2026} demonstrated GFM-BESS load smoothing and supply continuity, assuming sufficient state-of-charge (SOC) headroom for short-duration studies and assigning sustained energy management to a higher-level controller. These studies do not explicitly address reserving bidirectional BESS power and energy to sustain approved grid supply to data centers under bounded, unannounced demand changes during native load restoration.

Maintaining approved grid supply to a data center requires onsite flexibility in both directions. When facility demand rises, onsite resources must increase net supply; when it falls, they must reduce supply or absorb surplus power. A BESS near its lower energy limit has limited discharge duration, while one near its upper limit has little room to absorb additional energy.

The problem addressed here is sustaining approved point-of-common-coupling (PCC) supply despite bounded, unannounced demand changes while restoring native loads. Processing delays complicate this decision because resource states continue to evolve before an approved command is applied. This paper (i) schedules native load pickup and grid supply to data centers through sequential MILPs, reserving bidirectional onsite power and energy flexibility; (ii) incorporates processing delays and rechecks continuous-time BESS power and energy feasibility using updated states before command application; and (iii) demonstrates, through independent phasor-domain simulations of a low-inertia restoration scenario, that frequency-aware load pickup can still violate operating limits when onsite resources cannot sustain approved grid supply to data centers.

\begin{figure}[!t]
\centering
\includegraphics[width=0.8\columnwidth]{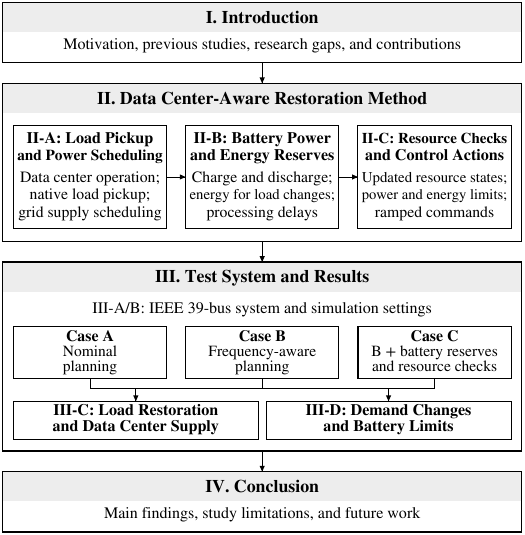}
\caption{Overall organization of the paper.}
\label{fig:overview2}
\end{figure}

Fig.~\ref{fig:overview2} shows the overall organization of the paper.
\section{Data Center-Aware Restoration Method}
\label{sec:method}\label{sec:model}
\subsection{Load Pickup and Power Scheduling}
Indices $j,i,g$ denote native load tasks, data-center facilities, and grid sources. Native demand is $P_{\rm N}=P_{\rm base}+\sum_j(n_j+c_j)$, where $n_j$ is additional restored service and $c_j$ is cold-load excess. Reactive demand follows each bus's nominal $Q/P$ ratio. Facility power balance is
\begin{equation}
q_i^a+D_i+B_i+F_i=d_i-U_i,\qquad L_i=q_i^a-q_i,
\label{eq:local}
\end{equation}
where $q_i^a,q_i$ are actual and approved grid supply at the PCC; $D_i,B_i,F_i$ are diesel, BESS, and fast-buffer powers; and $d_i$ is demand. Positive storage power denotes discharge; $U_i>0$ denotes unmet demand and $U_i<0$ unabsorbed excess. Exporting power from facility to the grid is prohibited in this work, and reactive import follows $\cos\varphi_i$. Local dynamics include
\begin{align}
\dot D_i&=\operatorname{clip}\!\left((u_i-D_i)/T_{D,i};-r_{D,i},r_{D,i}\right),\label{eq:diesel}\\
\dot E_i&=-\Psi_{\eta_{B,i}}(B_i)/\kappa_t,\quad
\dot e_i=-\Psi_{\eta_{F,i}}(F_i),\label{eq:storage}
\end{align}
Here $u_i,T_{D,i},r_{D,i}$ are diesel command, lag, and slew, and $\Psi_\eta(p)=\max\{p/\eta,\eta p\}$ for efficiency $0<\eta\le1$. Time is in seconds, $E_i$ in MWh, $e_i$ in MJ, and $\kappa_t=3600$~s/h. Resources have finite response and power/energy limits.

Requests at $t_m$, spaced by $\Delta$, use observed states, remaining ramps, topology, commitments, and demand reference $d_i^{\rm ref}(t_m)$ from current/past measurements. Operators declare bounds $d_i^{\rm lo}\le d_i(t)\le d_i^{\rm hi}$ from workload knowledge; these remain fixed in this study. For each candidate action set satisfying all prerequisites $\mu\in\mathcal M_m$, including the empty batch, solve
\begin{equation}
\begin{aligned}
J_m^\star(\mu)=\min_{\hat{\boldsymbol z}\in\hat{\mathcal F}_m(\mu)}
\biggl\{&\Delta_h\sum_{\ell=1}^{N_p}\mathcal L_\ell\\
&+\kappa_s\sum_j\xi_j+\kappa_\mu\nu(\mu)\biggr\},
\end{aligned}
\label{eq:objective}
\end{equation}
Here $\hat{\boldsymbol z}$ collects predicted decisions, $\hat{\mathcal F}_m$ is their feasible set, $\ell$ indexes intervals, and $\Delta_h=\Delta/\kappa_t$. Priority $\pi_j$ rewards service; $\xi_j$ denotes deadline shortfall and $\nu(\mu)$ counts actions.
The running cost is
\begin{equation}
\mathcal L_\ell=-c_N\sum_j\pi_j\hat n_{\ell j}-c_q\sum_i\hat q_{\ell i}+C_\ell.
\label{eq:running_cost}
\end{equation}
Cost $C_\ell$ weights BESS discharge/charge, diesel commands, and source output by $c_{\rm dis},c_{\rm ch},c_D,c_g$.

The MILP includes standard constraints used in power system optimization; only selected equations are presented here for brevity. The MILP enforces nondecreasing load pickup blocks $\hat n_{\ell j}=n_j^{\rm pkt}x_{\ell j}$, $x_{\ell j}\in\mathbb Z_{\ge0}$; release/connectivity rules; cold-load pickup; resource limits, slews, and energy; diesel lag; branch-flow limits; and heuristic frequency/RoCoF screens. Each stage $T_{\rm st}$ permits one positive pickup commitment up to $\bar N_{\rm st}$. The buffer has no sustained schedule.
Nominal generation covers native demand, grid supply at the PCC, and a loss allowance $\epsilon_{\rm loss}$. Facility balance is $\hat q_{\ell i}+\hat D_{\ell i}+\hat b^+_{\ell i}-\hat b^-_{\ell i}=d_i^{\rm ref}(t_m)$; nonnegative BESS discharge/charge powers update energy by $\Delta\hat E_{\ell i}=-\Delta_h(\hat b^+_{\ell i}/\eta_{B,i}-\eta_{B,i}\hat b^-_{\ell i})$. Simultaneous flows are relaxed and penalized.

\subsection{Battery Power and Energy Reserves}

With BESS tracking reserve $\epsilon_{P,i}$, usable power is $R_i=\bar B_i-\epsilon_{P,i}$. For any declared demand, grid--diesel supply must lie between $s_i^{\rm lo}=d_i^{\rm hi}-R_i$ and $s_i^{\rm hi}=d_i^{\rm lo}+R_i$, requiring $d_i^{\rm hi}-d_i^{\rm lo}\le2R_i$.
Within battery reserve horizon $T_{\rm ready}$, predicted supply at endpoints $\ell\Delta$ and the steady-state supply under the first control setpoints satisfy
\begin{equation}
\begin{aligned}
s_i^{\rm lo}&\le\hat q_{\ell i}+\hat D_{\ell i}
\le s_i^{\rm hi},\\
s_i^{\rm lo}&\le\hat q_{1i}+\hat u_{1i}
\le s_i^{\rm hi}.
\end{aligned}
\label{eq:prefix_power}
\end{equation}
Both supplies are also constrained to the demand
interval $[d_i^{\rm lo},d_i^{\rm hi}]$.

To ensure feasibility between endpoints, the grid--diesel supply during the first ramp
$0\le\sigma\le T_r=\Delta$ is modeled as
\begin{equation}
\begin{aligned}
&s_i^{(1)}(\sigma)=(1-\alpha)q_{0i}
+\phi_iD_{0i}+(1-\phi_i-\omega_i)u_{0i}\\
&\quad+\alpha\hat q_{1i}+\omega_i\hat u_{1i},\qquad
\alpha=\sigma/T_r,\\
&\phi_i=e^{-\sigma/T_{D,i}},\quad
\omega_i=[\sigma-T_{D,i}(1-\phi_i)]/T_r.
\end{aligned}
\label{eq:first_supply}
\end{equation}
Here $(q_{0i},u_{0i},D_{0i})
=(q_i,u_i,D_i)(t_m)$ are the states at the current request time.
At samples $\sigma_k=kT_s$,
$k=1,\ldots,T_r/T_s$, impose
\begin{equation}
s_i^{\rm lo}+\epsilon_i^{\rm samp}
\le s_i^{(1)}(\sigma_k)
\le s_i^{\rm hi}-\epsilon_i^{\rm samp},
\label{eq:samples}
\end{equation}
Here $\epsilon_i^{\rm samp}=\Gamma_iT_s^2/8+\epsilon_{{\rm num},i}$ bounds interpolation and numerical errors. For the unclipped diesel model, $|\ddot s_i^{(1)}|\le\Gamma_i$ (MW/s$^2$), with
\begin{equation}
\Gamma_i=\frac{|D_{0i}-u_{0i}|}{T_{D,i}^2}
+\frac{\max\{u_{0i},\bar u_i-u_{0i}\}}{T_rT_{D,i}},
\label{eq:curvature}
\end{equation}
Here $\bar u_i$ is the diesel-command limit and $\epsilon_{{\rm num},i}=10^{-4}$~MW.

During processing delay $\delta\in[0,T_d]$,
previous ramps continue and hold, yielding
retained states
$q_i^r(\delta),u_i^r(\delta),D_i^r(\delta)$.
These replace the initial states in
\eqref{eq:first_supply}; linear cuts enforce
\eqref{eq:samples} across the resulting
delay-dependent envelopes, including diesel
prediction error.
Let $\tilde s_i(\sigma;\delta)$ denote the
delayed first ramp followed by fixed
$\hat q_{1i},\hat u_{1i}$ and diesel lag. Its cumulative supply energy is
$W_i(T;\delta)=\kappa_t^{-1}
\int_0^T\tilde s_i(\sigma;\delta)\,d\sigma$.
To prevent BESS depletion or overcharging under
extreme demand, at $T=\ell\Delta\le T_{\rm ready}$
impose, with energy margin
$\epsilon_E^{\rm plan}$,
\begin{equation}
\begin{aligned}
W_i(T;\delta)\ge{}&
\frac{d_i^{\rm hi}T+\bar B_i\delta}{\kappa_t}
-\eta_{B,i}[E_i(t_m)-E_i^{\min}
-\epsilon_E^{\rm plan}],\\
W_i(T;\delta)\le{}&
\frac{d_i^{\rm lo}T-\bar B_i\delta}{\kappa_t}
+\frac{E_i^{\max}-E_i(t_m)
-\epsilon_E^{\rm plan}}{\eta_{B,i}}.
\end{aligned}
\label{eq:energy_screens}
\end{equation}
The $\bar B_i\delta$ terms reserve processing-time energy movement. Retained-segment extrema give finite linear constraints. With consistent states/margins and supply within demand bounds, these test endpoint energy, not intermediate-time feasibility.

\subsection{Resource Checks and Control Actions}
At application $t_a$, use fresh states and supply $s_i(\sigma)=q_i(t_a+\sigma)+D_i(t_a+\sigma)$. For $d\in\{d_i^{\rm lo},d_i^{\rm hi}\}$, ideal BESS continuation is
\begin{equation}
E_i^{(d)}(t_p)=E_i(t_a)-\frac{1}{\kappa_t}\int_0^{t_p}
\Psi_{\eta_{B,i}}\!\bigl(d-s_i(\sigma)\bigr)\,d\sigma.
\label{eq:exact_energy}
\end{equation}
For every $0\le t_p\le T_{\rm res}=\min\{T_{\rm ready},T_{\rm obs}-t_a\}$, require
\begin{IEEEeqnarray}{c}
s_i^{\rm lo}\le s_i(t_p)\le s_i^{\rm hi},\qquad
|\dot D_i(t_a+t_p)|\le r_{D,i},\IEEEnonumber\\
E_i^{\min}+\epsilon_E\le E_i^{(d_i^{\rm hi})}(t_p),\qquad
E_i^{(d_i^{\rm lo})}(t_p)\le E_i^{\max}-\epsilon_E,\IEEEnonumber\\
e_i^{\min}+\epsilon_F\le e_i(t_a)\le e_i^{\max}-\epsilon_F.
\label{eq:certificate}
\end{IEEEeqnarray}
Extrema occur at segment boundaries, supply stationary points, and crossings $s_i=d$. Monotonicity of $\Psi_\eta$ bounds ideal energy for every demand trajectory within the declared bounds; the diesel-rate test verifies the predicted diesel response without active ramp limiting. The 	continuous-time BESS power and energy checks are conditional on fresh states, declared bounds, and fixed controls. The buffer row checks initial energy only. Its 0.5~MJ provides just 10~ms at 50~MW, or 5~ms from the initial midpoint, neglecting losses; sustained buffer feasibility is not verified.

In Algorithm~\ref{alg:restoration}, quasi-steady-state (QSS) checks solve droop-distributed alternating-current (AC) balances at 7.5-s ramp samples and pickup breakpoints, testing voltage, current, branch loading, and synchronous-generator (SG) reactive power. B/C also require frequency deviation within 0.45~Hz. An aggregate swing observer, separate from phasor replay, uses summed inertia, source droop/lag/limits, actual, and a fixed initial loss fraction; B/C require its frequency within 0.08~Hz of 60~Hz at acceptance. Application at $t_a=t_m+\tau_m$ requires $\tau_m\le T_d$, final verification within $T_c$, and a passing post-delay audit. Commands ramp over $T_r$; pickup starts after $T_N$ at rate $r_N$. Changed controls/bounds end the previous verified battery operating interval; rejection grants no extension. Retained ramps require fresh checks.

\begin{algorithm}[!t]
\footnotesize
\caption{Data Center-Aware Restoration Procedure}
\label{alg:restoration}
\begin{algorithmic}[1]

\Require Current states, ongoing ramps, equipment availability,
and demand reference/bounds.

\State Form batches satisfying all action prerequisites;
bound equipment-state changes over the allowed processing-time
range $0 \leq \delta \leq T_d$.

\ForAll{batches, including the empty batch}
  \State Solve the MILP; repair AC feasibility; check QSS,
  the frequency observer, and battery power and energy limits.

  \State Retry eligible failures with less pickup.
  Failed battery checks permit no extra pickup.
  Keep the first passing proposal.
\EndFor

\State Rank proposals using the objective corrected for repaired
first-interval source cost; advance existing controls through
the search delay.

\ForAll{ranked proposals}
  \State Correct endpoints; recheck availability, AC/QSS,
  the frequency observer, and battery limits; advance existing
  controls through the acceptance delay.

  \If{updated checks pass}
    \State Verify the updated PCC ramp rate and battery limits;
    advance existing controls through the closing delay.

    \If{verification, deadlines, and the post-delay audit pass}
      \State Apply actions and ramp the first control endpoints
      over $T_r$; start eligible native load pickup after $T_N$
      at rate $r_N$.

      \State Preserve native ramps for equipment-only updates;
      end the previous verified operating interval, verify the
      new interval, and \Return.
    \EndIf
  \EndIf
\EndFor

\State Retain controls without new pickup; extend the verified
operating interval only after updated path, frequency-observer,
battery, timing, and post-delay checks pass.

\end{algorithmic}
\end{algorithm}

\section{Test System and Results}
\label{sec:case_study}
\subsection{IEEE 39-bus System and Simulation Settings}
\begin{figure}[!ht]
\centering
\includegraphics[width=0.6\columnwidth]{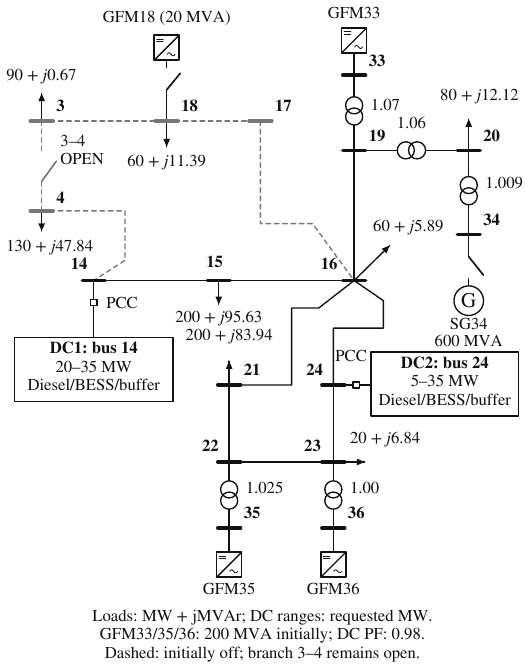}
\caption{Restoration island and prepared extensions. Dashed elements are initially de-energized; branch 3--4 remains open.}
\label{fig:sld}
\end{figure}

The modified IEEE 39-bus system starts with a 13-bus island, data centers at buses 14 and 24, and 200~MW of restored native load (Fig.~\ref{fig:sld}). Table~\ref{tab:restoration_inputs} lists an additional 640~MW of native load and its release times. SG34 becomes available at 2820~s. GFM33/GFM35/GFM36 initially have ratings of 200~MVA each. Rating upgrades to 300/300/250, 400/450/350, and 500/550/400~MVA become available at 2820, 3000, and 3600~s, respectively. Each action requires its prerequisites to be met and must pass checks using updated operating states.

\begin{table}[!t]
\caption{Native Load Restoration Tasks}
\label{tab:restoration_inputs}
\centering\footnotesize
\setlength{\tabcolsep}{3pt}
\renewcommand{\arraystretch}{0.95}
\begin{tabular*}{\columnwidth}{@{\extracolsep{\fill}}lrrrrr@{}}
Task & Bus & Added MW & Release (s) & Deadline (s) & $\pi_j$\\\hline
B20 & 20 & 20 & 0 & 600 & 1.0\\
B18 & 18 & 60 & 600 & 1200 & 1.2\\
B3 & 3 & 90 & 1200 & 1800 & 1.0\\
B4 & 4 & 130 & 1800 & 2400 & 1.1\\
B15 & 15 & 160 & 0 & 3000 & 1.0\\
B21 & 21 & 180 & 0 & 3600 & 1.2\\\hline
\end{tabular*}
\par\smallskip
\parbox{\columnwidth}{\raggedright Added loads exclude initial service; deadlines are soft.}
\end{table}

Case A uses nominal planning, 20-MW/s pickup, GFM ceiling $0.90S_g$, and SG ceiling 508~MW. B adds conservative dispatch, stage caps, frequency/RoCoF screens, and tighter checks. C adds battery reserve scheduling, power and energy checks, and repeated checks for unchanged controls. B/C share grid margins and pickup rules, isolating the additional battery reserve constraints and checks; A also differs in ramp/dispatch limits. Hardware, initial states, objectives, workloads, and delays are common.

Demand is reconstructed from published profiles \cite{noauthor_large_nodate,nerc_large_loads_2025}. Added 30-s events drive bus 14 to 20~MW at 1800~s and bus 24 to 35~MW at 2700~s, with 0.1-s rise and fall times. The controller knows bounds, not future events. Table~\ref{tab:parameters} gives C settings. B/C need at least seven 600-s stages to add 640~MW under the 100-MW cap, precluding completion before 3600~s even without delays/ramps. Unused allowance cannot be recovered within a stage.

\begin{table}[!ht]
\caption{Principal Case C Parameters and Initial Conditions}
\label{tab:case_c_parameters}\label{tab:parameters}
\centering\footnotesize
\setlength{\tabcolsep}{3pt}
\renewcommand{\arraystretch}{0.95}
\begin{tabularx}{\columnwidth}{@{}>{\raggedright\arraybackslash}Xl@{}}
\hline
Parameter & Value\\\hline
Request/ramp, $\Delta,T_r$; stage/reserve & 60~s; 600~s\\
Horizons, $N_p\Delta/T_{\rm obs}$ & 1800/4800~s\\
Total/closing deadline, $T_d,T_c$ & 10/0.15~s\\
Pickup rate/onset, $r_N,T_N$ & 2~MW/s; 3~s\\
pickup block size/stage limit; cold-load fraction/decay & 10/100~MW; 0.15/120~s\\
Demand bounds, $[d_i^{\rm lo},d_i^{\rm hi}]$ & $[20,35]/[5,35]$~MW\\
Available BESS power, $\bar B_i$ & 15/18~MW\\
Installed BESS rating/energy & 50~MVA/10~MWh\\
Initial BESS energy; bounds & 5.5~MWh; $[2,9]$~MWh\\
Initial diesel; limit $\bar u_i$ & 30/20~MW; 35~MW\\
Diesel lag/slew & 10~s/1~MW/s\\
PCC cap/slew; power factor & 36.47952~MW/0.3~MW/s; 0.98\\
Buffer power/capacity/initial energy & 50~MW/0.5~MJ/0.25~MJ\\
BESS/buffer efficiency & 0.97/0.99\\
BESS nominal delay/lag & 2/1~ms\\
BESS energy margins, $\epsilon_E^{\rm plan}/\epsilon_E$ & 0.04/0.03~MWh\\
Buffer guard, $\epsilon_F$; sample period $T_s$ & 0.12~MJ; 1~s\\
GFM/SG droop &  4/5\%\\
GFM/SG response lag; dispatch ceiling & 0.08/0.6~s; $0.85S_g$/450~MW\\
Grid-storage energy per rating; initial SOC & 2~MWh/MVA; 80\%\\
Planning loss/energy factor & 1\%/1.25\\\hline
\end{tabularx}
\par\smallskip
\parbox{\columnwidth}{\raggedright Facility pairs denote buses 14/24; $S_g$ is current GFM rating; PCC power factor is lagging.}
\end{table}

This is a low-inertia scenario; demand bounds and margins are study choices. The 2.25-MW reserve gives usable BESS powers 12.75/15.75~MW. Objective weights $(c_N,c_q,\kappa_s,\kappa_\mu)=(10,1,1000,0.02)$ favor service/deadlines; $(c_{\rm dis},c_{\rm ch},c_D)=(0.05,1.05,0.20)$ and $c_g=0.04/0.01$ penalize facility and GFM/SG output.

AC acceptance uses constant active/reactive-power loads. Replay uses native constant-impedance/current/power (ZIP) fractions, normalized to initial voltage, and fixed PCC power factor. Operating limits are 59.5--60.5~Hz, backward 100-ms RoCoF of 1.5~Hz/s, and 0.94--1.06~pu voltage.

\subsection{Implementation and Computation}
The study uses relative gap $10^{-5}$ and a 10,000-node limit. The largest Case C problem contains 1026 decision variables and 1891 constraints. Building and solving an individual MILP in Case C took a median of 0.041 s and a maximum of 0.122 s. Across 80 requests per case, the median and maximum times for search, acceptance, and closing were 3.38 and 18.06 s for A, 1.04 and 14.80 s for B, and 0.93 and 8.97 s for C. The corresponding numbers of requests that missed a computation deadline were 1, 5, and 0. These measured times exclude simulation state propagation, warmup, saving outputs, checks after the modeled delay, and independent dynamic replay. The final check at command application was timed separately and took at most 0.087 s in Case C.

\subsection{Load Restoration and Data Center Supply}
\begin{figure}[!t]
\centering
\includegraphics[width=\columnwidth]{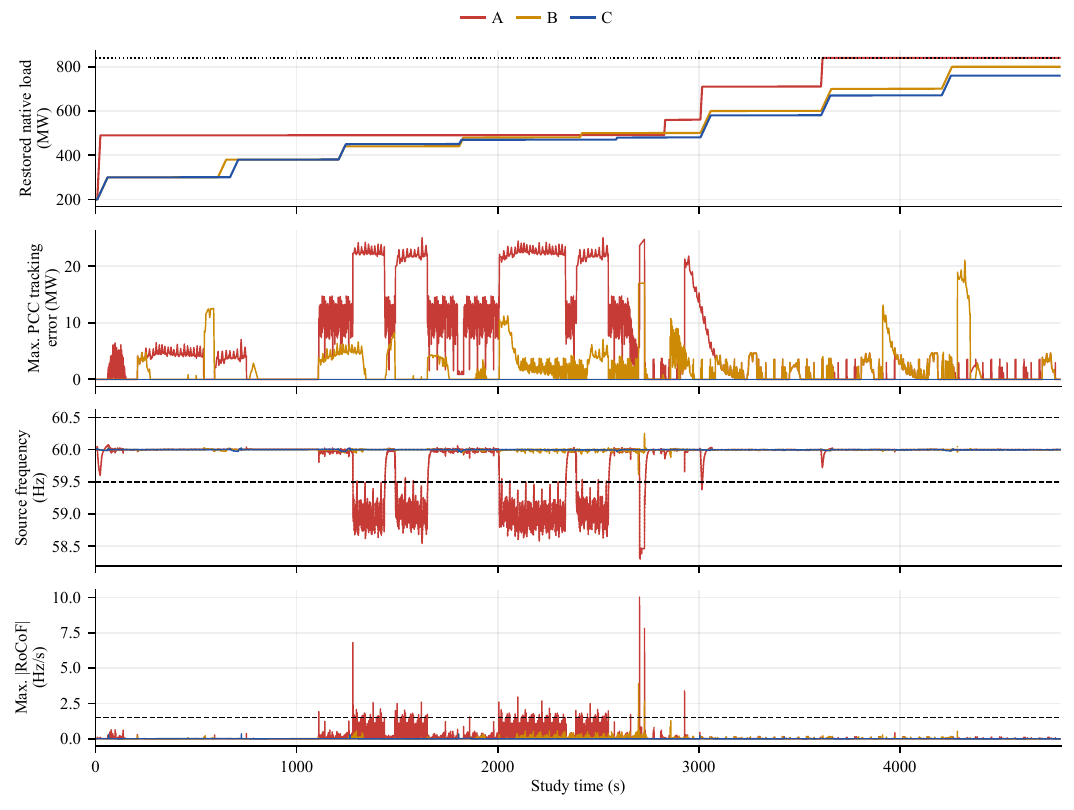}
\caption{A--C restoration and phasor replay: native service, maximum absolute PCC error, source-frequency envelope, and maximum absolute backward 100-ms RoCoF.}
\label{fig:overview}
\end{figure}

Fig.~\ref{fig:overview} and Table~\ref{tab:results} show that Case A violates frequency and RoCoF limits and B exceeds RoCoF while meeting frequency bounds. Both briefly violate the SG reactive lower bound. C passes all tested limits, leaving 80~MW at bus 15 unrestored. C picks up 20~MW less than B before 3000~s and 10~MW less in each of the last two stages. A/B continue diagnostically after violations; their larger totals are not secure outcomes.

\begin{table}[!ht]
\caption{Recorded Outcomes Over 4800 s}
\label{tab:results}
\centering\footnotesize
\setlength{\tabcolsep}{3pt}
\renewcommand{\arraystretch}{0.95}
\begin{tabularx}{\columnwidth}{@{}>{\raggedright\arraybackslash}Xrrr@{}}
\hline
Metric & A & B & C\\\hline
Final native service (MW) & 840 & 800 & 760\\
Native energy served (MWh) & 807.83 & 695.03 & 675.23\\
Peak absolute PCC error (MW) & 25.064 & 21.070 & 0.000\\
Minimum source frequency (Hz) & 58.3017 & 59.6110 & 59.9603\\
Peak 100-ms RoCoF (Hz/s) & 10.0209 & 3.9033 & 0.3484\\
Grid operating checks & Fail & Fail & Pass\\\hline
\end{tabularx}
\par\smallskip
\parbox{\columnwidth}{\raggedright Native service excludes grid supply at the PCC.}
\end{table}

\begin{figure}[!t]
\centering
\includegraphics[width=\columnwidth]{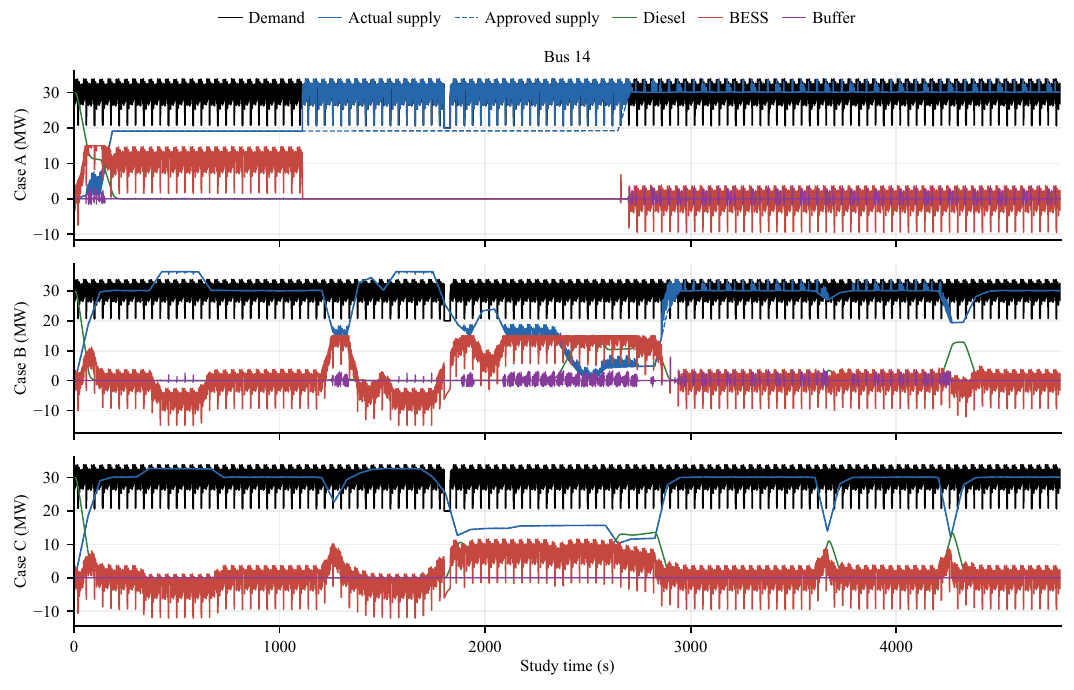}
\caption{Bus-14 demand, approved/actual grid supply at the PCC, diesel, BESS, and buffer powers for A--C. Storage power is positive on discharge.}
\label{fig:facility14}
\end{figure}

\begin{figure}[!t]
\centering
\includegraphics[width=\columnwidth]{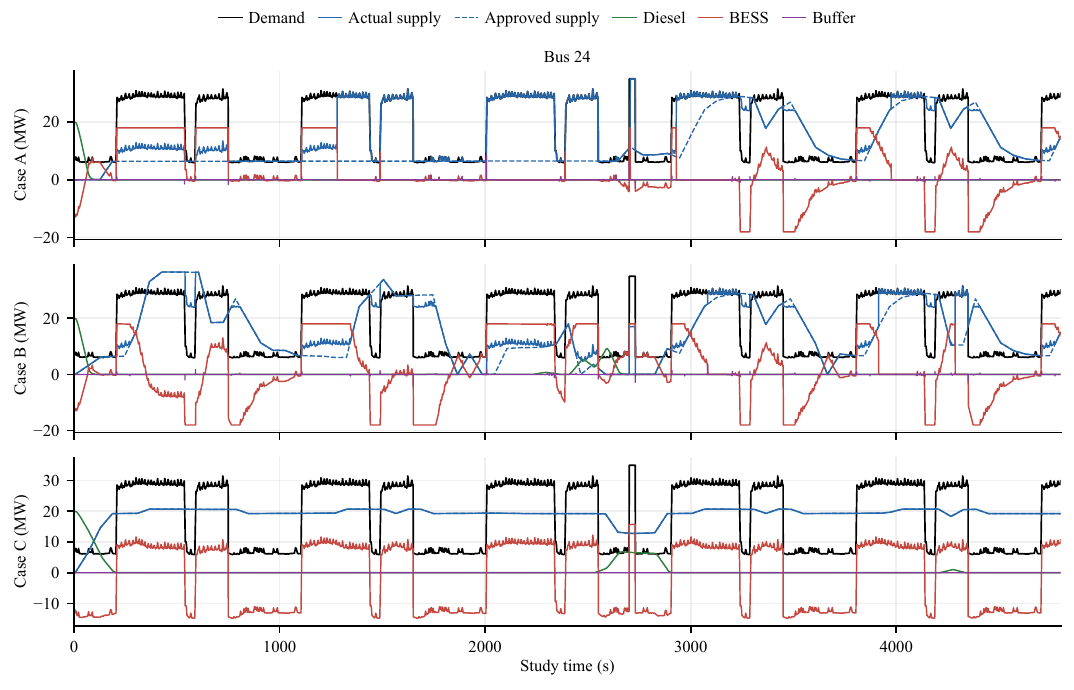}
\caption{Bus-24 power sharing for A--C, using the signals and sign convention of Fig.~\ref{fig:facility14}.}
\label{fig:facility24}
\end{figure}

Figures~\ref{fig:facility14} and~\ref{fig:facility24} show A/B reaching the 2-MWh BESS floor and deviating from PCC commands. All cases have zero deliberate curtailment and numerically negligible unmet facility energy; A/B rely on unplanned grid draw. In Case C, the modeled buffer compensates for power mismatches, keeping PCC power close to its commanded value while remaining within the buffer’s operating limits.

\subsection{Demand Changes and Battery Limits}
\begin{figure}[!t]
\centering
\includegraphics[width=\columnwidth]{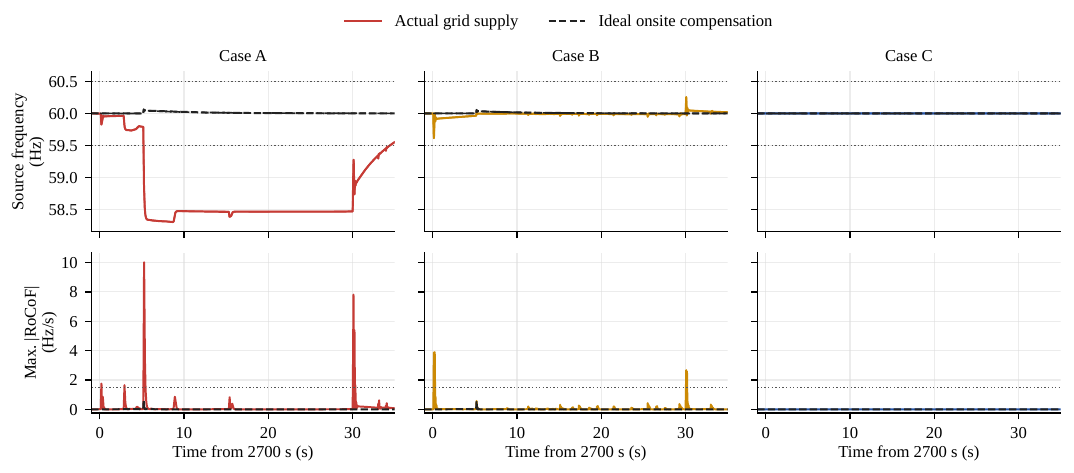}
\caption{Actual and ideal-compensated 2700-s event replay: frequency envelopes (top) and maximum absolute backward 100-ms RoCoF (bottom).}
\label{fig:causal}
\end{figure}

\begin{figure}[!t]
\centering
\includegraphics[width=\columnwidth]{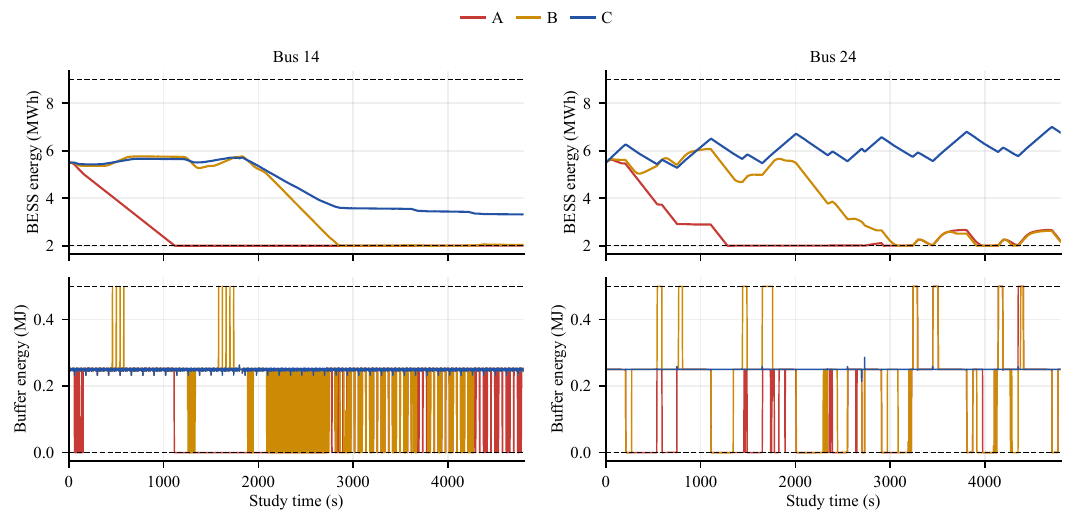}
\caption{Facility BESS/buffer energies.}
\label{fig:energy}
\end{figure}

Actual and ideal-compensated replay simulations start at 2699~s with identical voltages, source states, energies, and RoCoF history; commands, pickup, and switching stay fixed. An ideal onsite compensation removes subsequent PCC-error variation, retaining the initial offset to avoid an artificial step.
Ideal compensation reduces event-window peak RoCoF from 10.0209 to 0.6633~Hz/s in A and from 3.9033 to 0.5648~Hz/s in B; both operating-limit checks during the event window change to pass (Fig.~\ref{fig:causal}). C gives 0.0063~Hz/s in both branches. The comparison isolates PCC-error variation in this event window.
C's recorded BESS minima, 3.316/5.274~MWh, exceed the 2-MWh floor (Fig.~\ref{fig:energy}). Buffer/voltage ranges are 0.214--0.286~MJ/0.963--1.042~pu; peak current is 0.918~pu and source-MVA/branch ratios are 0.948/0.851.

\section{Conclusion}

The proposed restoration method satisfied all evaluated operating limits, whereas both benchmark policies violated those limits despite restoring more native load. These results indicate that secure restoration depends not only on generation availability but also on the ability of data centers to sustain approved grid supply at the PCC under changing demand. The lower pickup in Case C is associated with its additional battery-reserve constraints and resource checks under the adopted stage rules; this scenario does not establish a general restoration limit. The battery power and energy feasibility results remain conditional on declared demand bounds and assumptions of resource response. Future work should evaluate finite-bandwidth resource dynamics and broader demand and restoration scenarios to establish the approach's applicability beyond the conditions studied.
\\
\\
\\
\\
\bibliographystyle{IEEEtran}
\bibliography{references}
\end{document}